# Pressure-induced Insulator to Metal Transition of Mixed Valence Compound Ce(O,F)SbS$_2$

Ryo Matsumoto[a,b], Masanori Nagao[c], Masayuki Ochi[d], Hiromi Tanaka[e], Hiroshi Hara[a,b], Shintaro Adachi[a], Kazuki Nakamura[e], Ryo Murakami[e], Sayaka Yamamoto[e], Tetsuo Irifune[f], Hiroyuki Takeya[a], Isao Tanaka[c], Kazuhiko Kuroki[d], and Yoshihiko Takano[a,b]

[a]*National Institute for Materials Science, 1-2-1 Sengen, Tsukuba, Ibaraki 305-0047, Japan*
[b]*University of Tsukuba, 1-1-1 Tennodai, Tsukuba, Ibaraki 305-8577, Japan*
[c]*University of Yamanashi, 7-32 Miyamae, Kofu, Yamanashi 400-8511, Japan*
[d] *Department of Physics, Osaka University, Machikaneyama-cho, Toyonaka, Osaka 560-0043, Japan*
[e]*National Institute of Technology, Yonago College, 4448 Hikona, Yonago, Tottori 683-8502, Japan*
[f]*Geodynamics Research Center, Ehime University, Matsuyama, Ehime 790-8577, Japan*

**Abstract**

Transport properties of CeO$_{0.85}$F$_{0.15}$SbS$_2$ and undoped CeOSbS$_2$ under high pressure were investigated experimentally and theoretically. Electrical resistivity measurements of the CeO$_{0.85}$F$_{0.15}$SbS$_2$ single crystals were performed under various high pressures using a diamond anvil cell with boron-doped diamond electrodes. The samples showed the insulator to metal transition by applying high pressure up to 30-40 GPa. On the other hand, the undoped CeOSbS$_2$ showed almost same transport property with the F-doped sample under high pressure. The valence state analysis using X-ray photoelectron spectroscopy revealed a simple valence state of Ce$^{3+}$ in CeO$_{0.85}$F$_{0.15}$SbS$_2$ and mixed valence state between Ce$^{3+}$ and Ce$^{4+}$ in undoped CeOSbS$_2$. The valence fluctuation in Ce carried out the comparable transport nature in the both samples. A band calculation suggests that the undoped CeOSbS$_2$ could be metallic under high pressure of 30 GPa in accordance with the experimental results. A superior thermoelectric property of power factor in CeOSbS$_2$ was estimated under high pressure around 20 GPa in comparison with that of ambient pressure.

# 1. Introduction

After a discovery of $BiS_2$-layered superconductors of $Bi_4O_4S_3$ [1], explorations for related superconducting compounds are accelerated rapidly, such as $R(O,F)BiS_2$ ($R$: La, Ce, Pr, Nd, Yb) [2-6]. These perspirations for materials researches provide a knowledge that the $BiS_2$-based superconductors has large selectivity in terms of the construction of crystal structure. These compounds have a layered structure that is composed of conducting layers ($BiS_2$) and charge reservoir blocking layers. We can modify the combination of various conducting layers and blocking layers flexibly depending on a desired functionality in the material [7].

The superconductivity occurs in the conducting layer with carrier-doped blocking layer by elemental substitution generally, for example, F-substitution into the O site in the $CeOBiS_2$ [8,9]. Recently, the superconductivity was also reported in "F-free" $CeOBiS_2$ single crystal beyond expectation [10]. The valence state analysis by a core level X-ray photoelectron spectroscopy (XPS) revealed that the blocking layer of CeO provides the carrier for the superconductivity originated from a valence fluctuation between $Ce^{3+}$ and $Ce^{4+}$. The excess carrier from the valence fluctuation induces the superconductivity in $CeOBiS_2$ compound instead of general F-substitution carrier doping. The superconducting transition temperature in $CeOBiS_2$ is drastically increased by applying pressure because the corrugation in the $BiS_2$ layer would be flatter than that is in ambient pressure [10].

It is also recently focused on a superior thermoelectric performance in the $BiS_2$-layered compound as well as the superconductivity [11-13]. A remarkable dimensionless figure-of-merit (ZT) = 0.36 was particularly reported in hot-pressed LaOBiSSe [11]. According to a theoretical suggestion by Ochi et al., the thermoelectric performance of a power factor could be enhanced by Sb-substitution in Bi site due to the suppression of a spin-orbital coupling (SOC) originated from Bi atom and the enhanced low dimensionality of the electronic structure [14]. Both the weakened SOC and the enhanced low dimensionality in $SbS_2$-based compounds are expected to increase the density of states near the band edge. In general, the van-Hove singularity (vHs) of the band edge near the Fermi energy carries out the increase of the thermoelectric property of power factor [15,16]. The $Ce(O,F)SbS_2$ is appropriate material of $SbS_2$-based compounds because the carrier concentration in the blocking layer can be controlled by not only F-doping but also the valence fluctuation.

Although $Ce(O,F)SbS_2$ attracts interest in the superconductivity and thermoelectric performance, the synthesized single crystal was insulator in spite of the same blocking layer of $Ce(O,F)BiS_2$ [17]. Moreover, the undoped $CeOSbS_2$ shows the valence fluctuation of $Ce^{3+}$ and $Ce^{4+}$ as same as $CeOBiS_2$ superconducting compound. In this study, we examined a metallization of $Ce(O,F)SbS_2$ by using a high pressure application from a viewpoint of the superconductivity and thermoelectric performance. If the band edge of $Ce(O,F)SbS_2$ approaches to the Fermi level, the thermoelectric property would be maximum. If it crosses the Fermi level, the appearance of superconductivity is expected. To confirm aforementioned scenario, the $Ce(O,F)SbS_2$ were synthesized in single crystals. The valence states of Ce atom in both samples were investigated by the core-level XPS analysis. The high pressure resistivity measurements were performed by using a diamond anvil cell (DAC) with metallic diamond electrodes [18]. Moreover, we calculated a band structure and thermoelectric property of power factor under high pressure.

**2. Experimental and calculational procedures**

Single crystals of $CeO_{0.85}F_{0.15}SbS_2$ and undoped $CeOSbS_2$ were grown by similar way to the previous studies using an alkali metal flux method [19-21]. The valence states of obtained samples were estimated by XPS analysis using AXIS-ULTRA DLD (Shimadzu/Kratos) with AlK$\alpha$ X-ray radiation ($hv$ = 1486.6 eV), operating under a pressure of the order of $10^{-9}$ Torr. The samples were cleaved using scotch tape in a highly vacuumed pre-chamber in the order of $10^{-7}$ Torr. The background signals were subtracted by active Shirley method using COMPRO software [22]. The photoelectron peaks were analyzed by the pseudo-Voigt functions peak fitting. Resistivity measurements of $CeO_{0.85}F_{0.15}SbS_2$ and undoped $CeOSbS_2$ single crystals under high pressure were performed using an originally designed DAC with boron-doped diamond electrodes [23,24] on the bottom anvil of nanopolycrystalline diamond [25], as shown in fig.1. The sample was placed on the boron-doped diamond electrodes in the center of the bottom anvil. The surface of the bottom anvil except for the sample space and electrical terminal were covered by the undoped diamond insulating layer. The cubic boron nitride powders with ruby manometer were used as a pressure-transmitting medium. The applied pressure values were estimated by the fluorescence from ruby powders [26] and the Raman spectrum from the culet of top diamond anvil [27] by an inVia Raman Microscope (RENISHAW). The resistance was measured by a standard four-terminal method using a physical property measurement system (Quantum Design: PPMS).

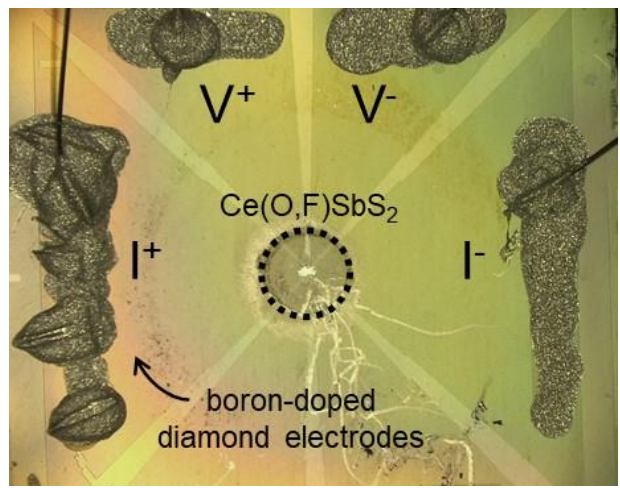

**Figure 1. Optical image of the sample space of DAC with boron-doped diamond electrodes**.

In our first principles calculations, a crystal structure was optimized under hydrostatic pressure using the projector augmented wave method [28] and the PBEsol exchange correlation functional [29] as implemented in the VASP code [30-33]. We employed a plane wave cutoff energy of 550 eV and a 12×12×4 $k$-mesh. We assumed a space group $P2_1/m$ as reported in an experimental study for CeOSbS$_2$ at ambient pressure [17]. After the structural optimization, we performed band structure calculation using the modified Becke-Johnson (mBJ) potential proposed by Tran and Blaha [34,35] and the full potential linearized augmented plane wave method as implemented in the WIEN2k code [36], in order to obtain a reliable size of the band gap. We employed an $RK_{max}$ parameter of 8.0 and a 14×14×4 $k$-mesh.

The thermoelectric properties were calculated using the BoltzTraP code [37] using the constant relaxation time approximation. For transport calculation, we used 200,000 $k$-points. For all the calculations, we included SOC and employed the open core treatment for Ce, which assumes the Ce$^{3+}$ valence state because accurate treatment of the $f$-electrons is very challenging for first principles calculations.

## 3. Results and discussion

XPS analysis was carried out to investigate the chemical states of Ce in CeO$_{0.85}$F$_{0.15}$SbS$_2$ and undoped CeOSbS$_2$ single crystals. The upper line in the Fig. 2 shows the Ce 3d core-level spectrum of CeOSbS$_2$ and result of the peak fitting by pseudo-Voigt functions referring the literatures [38,39]. The all peak positions of Ce$^{4+}$ and Ce$^{3+}$ were firstly determined by using those in standard materials of CeO$_2$ and Ce$_2$S$_3$, respectively. The main peaks at 880.7 eV and 885.2 eV corresponding to Ce 3d$_{5/2}$ with the valence state of Ce$^{3+}$. The peaks at 899.1 eV and 903.6 eV are attributed to Ce 3d$_{3/2}$ with the valence state of Ce$^{3+}$, from a consideration of spin–orbit splitting of 18.4 eV. Although the signal of the Ce 3d level has a very complicated satellite structure, it is clear that there is a single peak at 916.7 eV, and that has been reported that it is associated to the Ce 3d$_{3/2}$ and a characteristic feature of a presence of tetravalent Ce ions (Ce$^{4+}$) in Ce compounds, implying that the chemical state of Ce atoms in the single crystals is in a mixed valence state of Ce$^{4+}$ and Ce$^{3+}$. The area ratio of the fitted peaks of Ce$^{3+}$ to Ce$^{4+}$ was estimated to be about 1:0.48, suggesting the average valence of Ce is Ce$^{3.32+}$, which result indicates that the valence state in Ce is slightly electropositive compared to Ce$^{3+}$.

The XPS result of CeO$_{0.85}$F$_{0.15}$SbS$_2$ is also shown in the lower line of Fig. 2. When focusing on the binding energy around 916.7 eV, which is a peak position of fingerprint of Ce$^{4+}$, the Ce$^{4+}$ valence state seems to disappear after the F-doping. The average Ce valence was estimated to be Ce$^{3.05+}$ from this measurement. It indicates that the Ce valence is no longer fluctuate in the F-doped crystals. Accordingly, the total charge in the blocking layer in undoped CeOSbS$_2$ of +1.32 is higher than that in CeO$_{0.85}$F$_{0.15}$SbS$_2$ of +1.25.

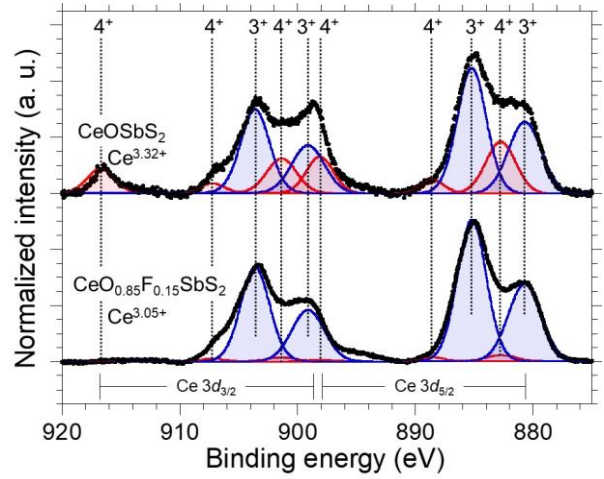

**Figure 2. Ce $3d_{5/2,3/2}$ XPS spectra and the fitted curves for $CeOSbS_2$ and $CeO_{0.85}F_{0.15}SbS_2$ single crystal.**

Figure 3 shows temperature dependence of resistance for (a) $CeO_{0.85}F_{0.15}SbS_2$ and (b) $CeOSbS_2$ single crystal under various pressures from 12 GPa to 55 GPa. $CeO_{0.85}F_{0.15}SbS_2$ behaved as an insulator with quite high resistance above 40 MΩ below 3 GPa. The resistance was decreased by applied pressure less than 100 kΩ at 12.5 GPa, and then we started to measure the electrical transport property of the sample. The sample firstly showed insulating behavior with band gap of 262.7 meV at 12.5 GPa. The resistance and band gap were continued to decrease with increase of the applied pressure. It suggests that the carrier transfer was started from the blocking layer $CeO_{0.85}F_{0.15}$ to the conducting layer $SbS_2$. The slope of resistance versus temperature ($dR/dT$) became negative at 37.4 GPa, namely a pressure-induced insulator to metal transition was occurred in $CeO_{0.85}F_{0.15}SbS_2$ single crystal. In contrast the undoped $CeOSbS_2$ presented relatively smaller resistance even near ambient pressure with few MΩ, which indicates the band gap of the undoped sample is smaller than that of F-doped sample. This is consistent with the carrier concentration in their blocking layer. On the other hand, the high pressure effects in the undoped sample is not so different compared to the F-doped sample, as shown in fig. 3(b). Although the resistance in both samples were dramatically decreased and the behaviors changed to metallic under 36.9 GPa, there was no sign of superconductivity up to 55 GPa. This result indicates the pressure or carrier concentration were not enough to induce the superconductivity. On the other hand, we successfully adjusted the transport property from insulator to metal. It could be expected that the superior thermoelectric property around the pressure of insulator to metal transition.

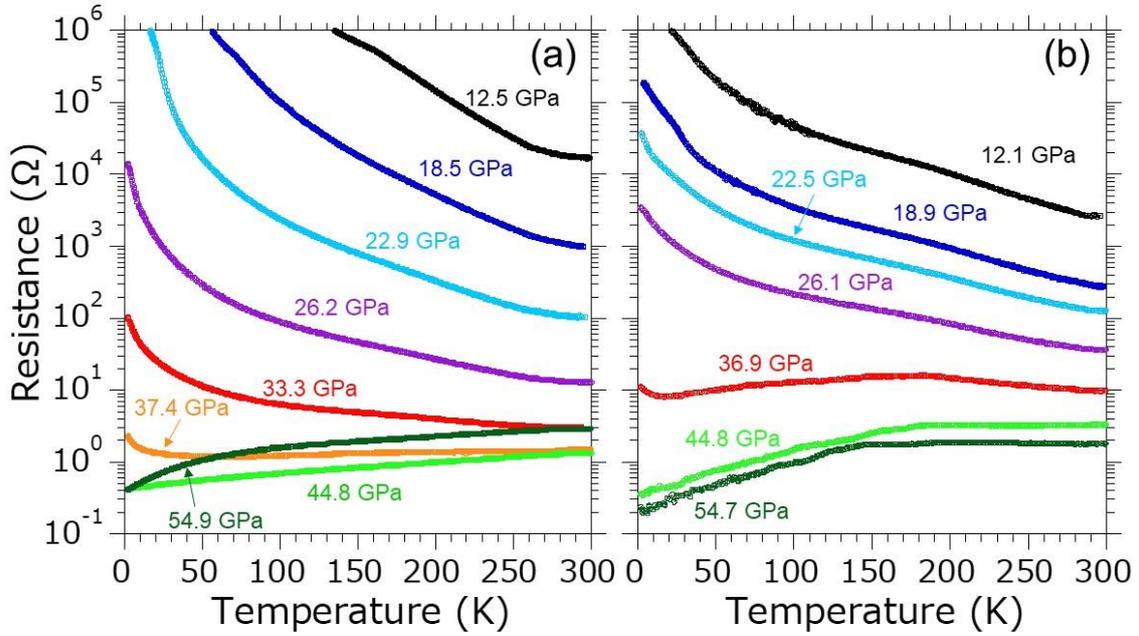

**Figure 3.** Temperature dependence of resistance in (a) CeO$_{0.85}$F$_{0.15}$SbS$_2$, (b) CeOSbS$_2$ single crystal under various pressure from 12 GPa to 55 GPa.

We theoretically considered the potential high thermoelectric property in the undoped CeOSbS$_2$ which has higher carrier concentration in the blocking layer. Figure 4 shows the calculated electronic band structure of CeOSbS$_2$ under various pressures from 0 to 50 GPa. The band structure presented that CeOSbS$_2$ becomes metallic over 30 GPa in correspondence with the experimental result. Band gap closing is caused by several reasons. One of the main reasons is an increased bandwidth of the conducting SbS$_2$ layer by applying pressure. Another one is an enhanced bilayer coupling between the SbS$_2$ layers, which can be inferred from a large splitting of the conduction band bottom near the (π,0,0) point [40,41]. We can also see that carrier is moved from the valence band top near the Γ point, which consists of the electronic states in the blocking layer (e.g., see Ref. [42]), to the conduction band bottom near the (π,0,0) point, which consists of the electronic states in the conducting layers. While our theoretical treatment does not consider the Ce valence fluctuation, it is possible to say that the applied pressure enables the charge transfer from the blocking layer to conducting layer even without Ce valence fluctuation. Theoretical investigation of the effect of the valence fluctuation is an important but challenging future issue.

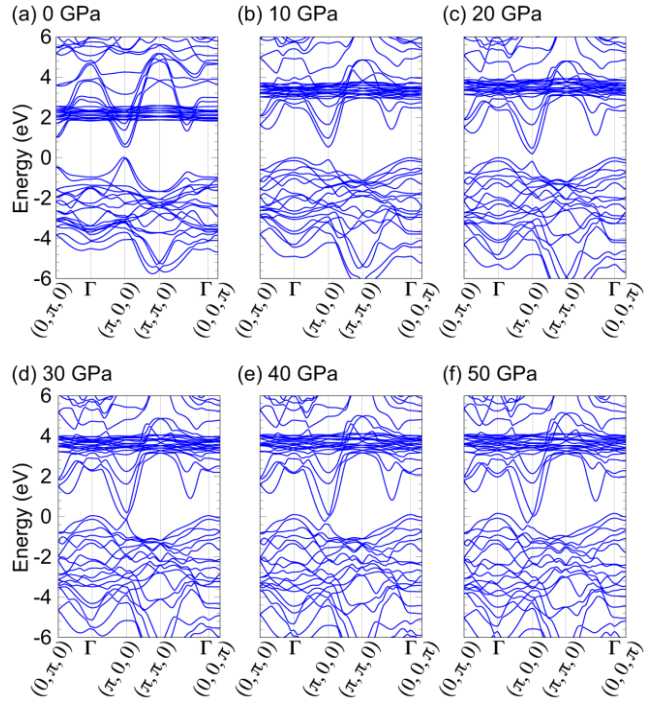

**Figure 4. Calculated electronic band structure of CeOSbS$_2$ under various pressures from 0 to 50 GPa.**

Figure 5 shows estimated power factor of CeOSbS$_2$ by our first principles calculation as a function of the electron carrier number under various pressures from 0 to 50 GPa. The calculation predicted that the maximum power factor could be obtained under 20 GPa for the *y* direction. Moreover, the superior power factor is realized in wide range of electron carrier number under high pressures. This is important thing to obtain real sample with the high thermoelectric property because exact carrier tuning for certain electron concentration is difficult. Because the band calculation as shown in fig. 4 was good corresponding to the experimental results, the theoretical calculations of thermoelectric property could be also considered reliable. Further attempt to realize such a high pressure state of this compound at ambient pressure is required, for example, high pressure annealing effect, lattice mismatch effect of thin film, and field effect-induced carrier doping.

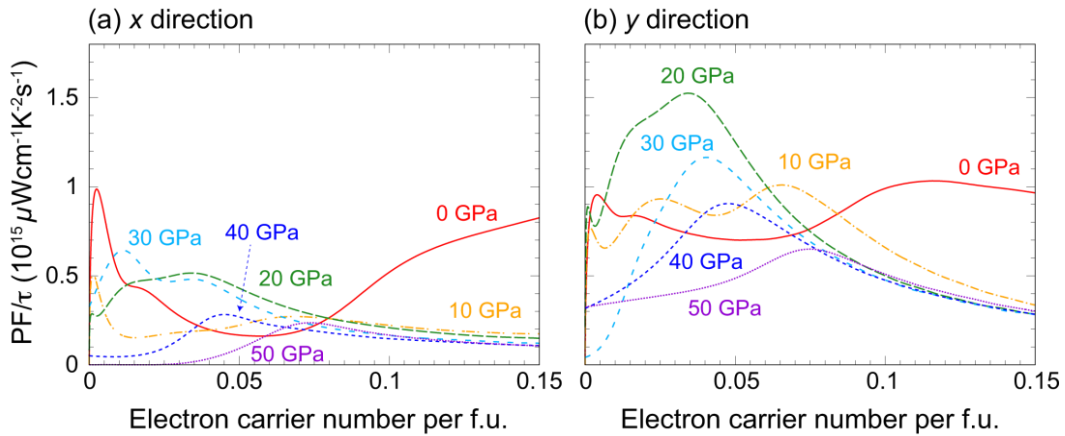

**Figure 5. Estimated power factor of CeOSbS$_2$ by the first principles calculation as a function of**

the electron carrier number under various pressures from 0 to 50 GPa. Power factor divided with the electron relaxation time τ for the (a) *x* and (b) *y* directions are shown, where the *y* axis is parallel to the $a_2$ direction and the *x* axis is orthogonal to the *y* axis in the conducting plane. For the *P2$_1$/m* space group, we set the $a_1$ and $a_3$ axes to be the non-orthogonal axes.

## 4. Conclusion

A potential superconductivity and thermoelectric property of $Ce(O,F)SbS_2$ was examined by theoretical and experimental approach using high pressure technique. $CeO_{0.85}F_{0.15}SbS_2$ and undoped $CeOSbS_2$ single crystals were used for the sample characterizations. The total carrier in blocking layer of undoped $CeOSbS_2$ was slightly higher than that of $CeO_{0.85}F_{0.15}SbS_2$ because of the mixed valence state in Ce atom. The samples showed the insulator-metal transition by applying high pressure up to 30-40 GPa. According to the calculation of electronic band structure under high pressure, the undoped $CeOSbS_2$ could be metallic under high pressure of 30 GPa in accordance with the experimental results. Our first principles calculation predicted the superior power factor around 20 GPa for the y direction than that at ambient pressure. Further attempt to realize such a high thermoelectric performance and unobserved superconductivity in this compound is required, for example, high pressure annealing effect, lattice mismatch effect of thin film, and field effect-induced carrier doping.

**Acknowledgment**

This work was partly supported by JST CREST Grant No. JPMJCR16Q6, JST-Mirai Program Grant Number JPMJMI17A2, JSPS KAKENHI Grant Number JP17J05926 and JP17H05481. A part of the fabrication process of diamond electrodes was supported by NIMS Nanofabrication Platform in Nanotechnology Platform Project sponsored by the Ministry of Education, Culture, Sports, Science and Technology (MEXT), Japan. The part of the high pressure experiments was supported by the Visiting Researcher's Program of Geodynamics Research Center, Ehime University.